\documentclass[aps,prd,preprint,preprintnumbers,tightenlines,superscriptaddress,showpacs,amsmath,amssymb]{revtex4}
\usepackage{graphicx} % Include figure files
\usepackage{dcolumn}  % Align table columns on decimal point
\usepackage{hyperref}
\hypersetup{colorlinks=true, linkcolor=blue}
\usepackage{color}

\graphicspath{{ps}}

\begin{document}

\vspace*{-3\baselineskip}
\resizebox{!}{3cm}{\includegraphics{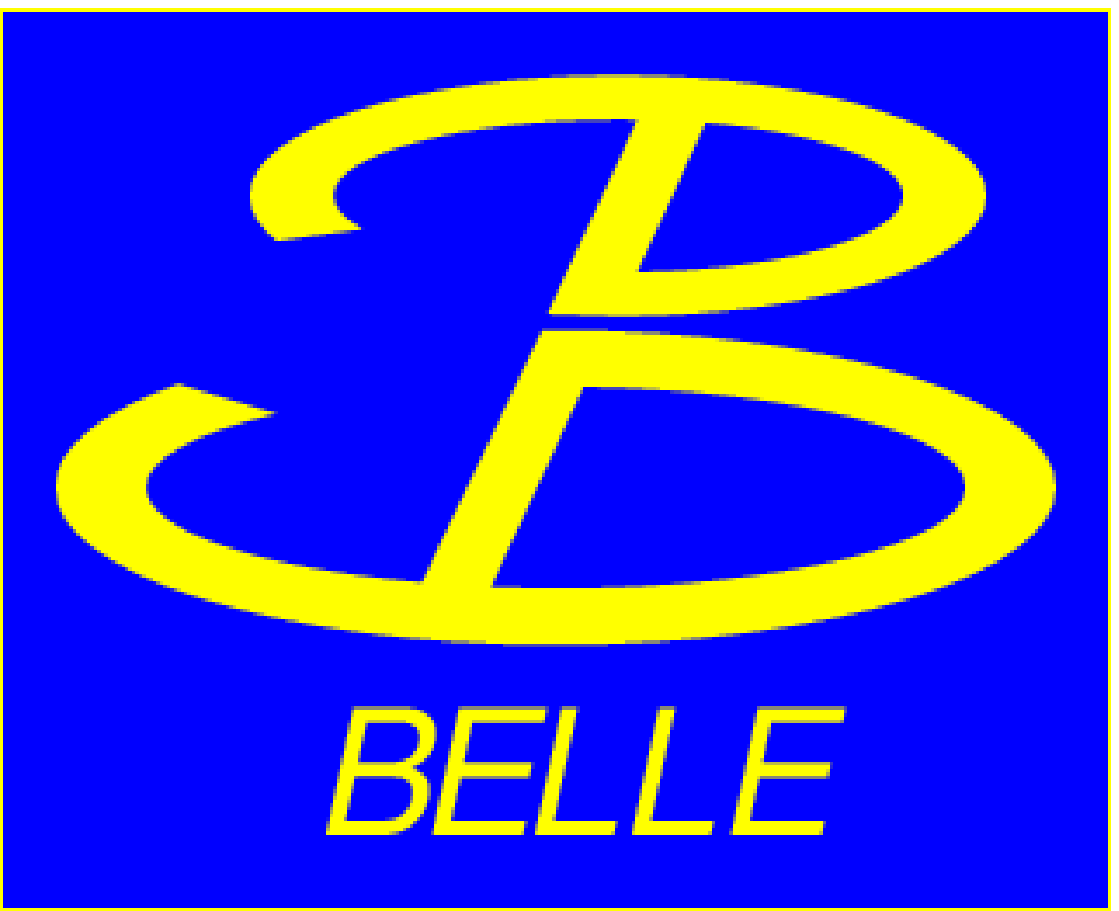}}

\preprint{\vbox{ \hbox{   }
			\hbox{Belle Preprint 2014-6}
		         \hbox{KEK Preprint 2013-67}}}

\title{ \quad\\[1.0cm] Measurements of the masses and widths of the $\boldsymbol{\Sigma_{c}(2455)^{0/++}}$ and $\boldsymbol{\Sigma_{c}(2520)^{0/++}}$ baryons}

\begin{abstract}
We present measurements of the masses and decay widths of the baryonic states $\Sigma_{c}(2455)^{0/++}$ and $\Sigma_{c}(2520)^{0/++}$ using a data sample corresponding to an integrated luminosity of 711 fb$^{-1}$ collected with the Belle detector at the KEKB $e^{+}e^{-}$ asymmetric-energy collider operating at the $\Upsilon(4S)$ resonance. We report the mass differences with respect to the $\Lambda_{c}^{+}$ baryon
\begin{eqnarray*}
M(\Sigma_{c}(2455)^{0})-M(\Lambda_{c}^{+}) & = & 167.29\pm0.01\pm0.02 \textrm{ MeV}/c^{2},\\
M(\Sigma_{c}(2455)^{++})-M(\Lambda_{c}^{+}) & = & 167.51\pm0.01\pm0.02 \textrm{ MeV}/c^{2},\\
M(\Sigma_{c}(2520)^{0})-M(\Lambda_{c}^{+}) & = & 231.98\pm0.11\pm0.04 \textrm{ MeV}/c^{2},\\
M(\Sigma_{c}(2520)^{++})-M(\Lambda_{c}^{+}) & = & 231.99\pm0.10\pm0.02 \textrm{ MeV}/c^{2},
\end{eqnarray*}
and the decay widths
\begin{eqnarray*}
\Gamma(\Sigma_{c}(2455)^{0}) & = & 1.76\pm0.04^{+0.09}_{-0.21} \textrm{ MeV}/c^{2},\\
\Gamma(\Sigma_{c}(2455)^{++}) & = & 1.84\pm0.04^{+0.07}_{-0.20} \textrm{ MeV}/c^{2},\\
\Gamma(\Sigma_{c}(2520)^{0}) & = & 15.41\pm0.41^{+0.20}_{-0.32} \textrm{ MeV}/c^{2},\\
\Gamma(\Sigma_{c}(2520)^{++}) & = & 14.77\pm0.25^{+0.18}_{-0.30} \textrm{ MeV}/c^{2},
\end{eqnarray*}
where the first uncertainties are statistical and the second are systematic. The isospin mass splittings are measured to be $M(\Sigma_{c}(2455)^{++})-M(\Sigma_{c}(2455)^{0})=0.22\pm0.01\pm0.01$ MeV/$c^{2}$ and $M(\Sigma_{c}(2520)^{++})-M(\Sigma_{c}(2520)^{0})=0.01\pm0.15\pm0.03$ MeV/$c^{2}$. These results are the most precise to date.
\end{abstract}


%%% Paper:    Sigma_c baryons
%%% Journal:  Physical Review D (Rapid Communication)
%%% Contacts: S. H. Lee (shlee@hep.korea.ac.kr)
%%%           E. Won (eunil@hep.korea.ac.kr)
%%%           B. R. Ko (brko@hep.korea.ac.kr)
%%% Non-responding authors or those who said NO are commented out.
%%% ====================================================================
%%% Click the RELOAD button on your web browser to see the updated file.
%%% ====================================================================
%%% Use \input{author} to insert this material into your latex file.
%%%%% Force institutions to appear in alphabetical order when typeset.
\noaffiliation
\affiliation{University of the Basque Country UPV/EHU, 48080 Bilbao}
\affiliation{Beihang University, Beijing 100191}
%%%\affiliation{University of Bonn, 53115 Bonn}
\affiliation{Budker Institute of Nuclear Physics SB RAS and Novosibirsk State University, Novosibirsk 630090}
\affiliation{Faculty of Mathematics and Physics, Charles University, 121 16 Prague}
%%%\affiliation{Chiba University, Chiba 263-8522}
\affiliation{University of Cincinnati, Cincinnati, Ohio 45221}
\affiliation{Deutsches Elektronen--Synchrotron, 22607 Hamburg}
%%%\affiliation{Department of Physics, Fu Jen Catholic University, Taipei 24205}
\affiliation{Justus-Liebig-Universit\"at Gie\ss{}en, 35392 Gie\ss{}en}
\affiliation{Gifu University, Gifu 501-1193}
\affiliation{II. Physikalisches Institut, Georg-August-Universit\"at G\"ottingen, 37073 G\"ottingen}
%%%\affiliation{The Graduate University for Advanced Studies, Hayama 240-0193}
\affiliation{Gyeongsang National University, Chinju 660-701}
\affiliation{Hanyang University, Seoul 133-791}
\affiliation{University of Hawaii, Honolulu, Hawaii 96822}
\affiliation{High Energy Accelerator Research Organization (KEK), Tsukuba 305-0801}
\affiliation{Hiroshima Institute of Technology, Hiroshima 731-5193}
\affiliation{IKERBASQUE, Basque Foundation for Science, 48011 Bilbao}
%%%\affiliation{University of Illinois at Urbana-Champaign, Urbana, Illinois 61801}
\affiliation{Indian Institute of Technology Guwahati, Assam 781039}
\affiliation{Indian Institute of Technology Madras, Chennai 600036}
%%%\affiliation{Indiana University, Bloomington, Indiana 47408}
\affiliation{Institute of High Energy Physics, Chinese Academy of Sciences, Beijing 100049}
\affiliation{Institute of High Energy Physics, Vienna 1050}
\affiliation{Institute for High Energy Physics, Protvino 142281}
%%%\affiliation{Institute of Mathematical Sciences, Chennai 600113}
\affiliation{INFN - Sezione di Torino, 10125 Torino}
\affiliation{Institute for Theoretical and Experimental Physics, Moscow 117218}
\affiliation{J. Stefan Institute, 1000 Ljubljana}
\affiliation{Kanagawa University, Yokohama 221-8686}
\affiliation{Institut f\"ur Experimentelle Kernphysik, Karlsruher Institut f\"ur Technologie, 76131 Karlsruhe}
\affiliation{Kavli Institute for the Physics and Mathematics of the Universe (WPI), University of Tokyo, Kashiwa 277-8583}
%%%\affiliation{Department of Physics, Faculty of Sciences, King Abdulaziz University, Jeddah 21589}
\affiliation{Korea Institute of Science and Technology Information, Daejeon 305-806}
\affiliation{Korea University, Seoul 136-713}
%%%\affiliation{Kyoto University, Kyoto 606-8502}
\affiliation{Kyungpook National University, Daegu 702-701}
\affiliation{\'Ecole Polytechnique F\'ed\'erale de Lausanne (EPFL), Lausanne 1015}
\affiliation{Faculty of Mathematics and Physics, University of Ljubljana, 1000 Ljubljana}
\affiliation{Luther College, Decorah, Iowa 52101}
\affiliation{University of Maribor, 2000 Maribor}
\affiliation{Max-Planck-Institut f\"ur Physik, 80805 M\"unchen}
\affiliation{School of Physics, University of Melbourne, Victoria 3010}
\affiliation{Moscow Physical Engineering Institute, Moscow 115409}
\affiliation{Moscow Institute of Physics and Technology, Moscow Region 141700}
\affiliation{Graduate School of Science, Nagoya University, Nagoya 464-8602}
\affiliation{Kobayashi-Maskawa Institute, Nagoya University, Nagoya 464-8602}
%%%\affiliation{Nara University of Education, Nara 630-8528}
\affiliation{Nara Women's University, Nara 630-8506}
\affiliation{National Central University, Chung-li 32054}
\affiliation{National United University, Miao Li 36003}
\affiliation{Department of Physics, National Taiwan University, Taipei 10617}
\affiliation{H. Niewodniczanski Institute of Nuclear Physics, Krakow 31-342}
\affiliation{Nippon Dental University, Niigata 951-8580}
\affiliation{Niigata University, Niigata 950-2181}
\affiliation{University of Nova Gorica, 5000 Nova Gorica}
\affiliation{Osaka City University, Osaka 558-8585}
%%%\affiliation{Osaka University, Osaka 565-0871}
\affiliation{Pacific Northwest National Laboratory, Richland, Washington 99352}
\affiliation{Panjab University, Chandigarh 160014}
\affiliation{Peking University, Beijing 100871}
%%%\affiliation{University of Pittsburgh, Pittsburgh, Pennsylvania 15260}
%%%\affiliation{Punjab Agricultural University, Ludhiana 141004}
%%%\affiliation{Research Center for Electron Photon Science, Tohoku University, Sendai 980-8578}
%%%\affiliation{Research Center for Nuclear Physics, Osaka University, Osaka 567-0047}
%%%\affiliation{RIKEN BNL Research Center, Upton, New York 11973}
%%%\affiliation{Saga University, Saga 840-8502}
\affiliation{University of Science and Technology of China, Hefei 230026}
\affiliation{Seoul National University, Seoul 151-742}
%%%\affiliation{Shinshu University, Nagano 390-8621}
\affiliation{Soongsil University, Seoul 156-743}
\affiliation{Sungkyunkwan University, Suwon 440-746}
\affiliation{School of Physics, University of Sydney, NSW 2006}
\affiliation{Department of Physics, Faculty of Science, University of Tabuk, Tabuk 71451}
\affiliation{Tata Institute of Fundamental Research, Mumbai 400005}
\affiliation{Excellence Cluster Universe, Technische Universit\"at M\"unchen, 85748 Garching}
\affiliation{Toho University, Funabashi 274-8510}
\affiliation{Tohoku Gakuin University, Tagajo 985-8537}
\affiliation{Tohoku University, Sendai 980-8578}
\affiliation{Department of Physics, University of Tokyo, Tokyo 113-0033}
\affiliation{Tokyo Institute of Technology, Tokyo 152-8550}
\affiliation{Tokyo Metropolitan University, Tokyo 192-0397}
\affiliation{Tokyo University of Agriculture and Technology, Tokyo 184-8588}
\affiliation{University of Torino, 10124 Torino}
%%%\affiliation{Toyama National College of Maritime Technology, Toyama 933-0293}
\affiliation{CNP, Virginia Polytechnic Institute and State University, Blacksburg, Virginia 24061}
\affiliation{Wayne State University, Detroit, Michigan 48202}
\affiliation{Yamagata University, Yamagata 990-8560}
\affiliation{Yonsei University, Seoul 120-749}
  \author{S.-H.~Lee}\affiliation{Korea University, Seoul 136-713} % Korea
  \author{B.~R.~Ko}\affiliation{Korea University, Seoul 136-713} % Korea
  \author{E.~Won}\affiliation{Korea University, Seoul 136-713} % Korea
  \author{A.~Abdesselam}\affiliation{Department of Physics, Faculty of Science, University of Tabuk, Tabuk 71451} % Tabuk
  \author{I.~Adachi}\affiliation{High Energy Accelerator Research Organization (KEK), Tsukuba 305-0801} % KEK
% \author{K.~Adamczyk}\affiliation{H. Niewodniczanski Institute of Nuclear Physics, Krakow 31-342} % Krakow
  \author{H.~Aihara}\affiliation{Department of Physics, University of Tokyo, Tokyo 113-0033} % Tokyo
% \author{S.~Al~Said}\affiliation{Department of Physics, Faculty of Science, University of Tabuk, Tabuk 71451}\affiliation{Department of Physics, Faculty of Science, King Abdulaziz University, Jeddah 21589) % Tabuk
% \author{K.~Arinstein}\affiliation{Budker Institute of Nuclear Physics SB RAS and Novosibirsk State University, Novosibirsk 630090} % BINP
% \author{Y.~Arita}\affiliation{Graduate School of Science, Nagoya University, Nagoya 464-8602} % Nagoya
  \author{D.~M.~Asner}\affiliation{Pacific Northwest National Laboratory, Richland, Washington 99352} % PNNL
% \author{T.~Aso}\affiliation{Toyama National College of Maritime Technology, Toyama 933-0293} % Toyama
% \author{V.~Aulchenko}\affiliation{Budker Institute of Nuclear Physics SB RAS and Novosibirsk State University, Novosibirsk 630090} % BINP
  \author{T.~Aushev}\affiliation{Institute for Theoretical and Experimental Physics, Moscow 117218} % ITEP
  \author{R.~Ayad}\affiliation{Department of Physics, Faculty of Science, University of Tabuk, Tabuk 71451} % Tabuk
% \author{T.~Aziz}\affiliation{Tata Institute of Fundamental Research, Mumbai 400005} % Tata
  \author{A.~M.~Bakich}\affiliation{School of Physics, University of Sydney, NSW 2006} % Sydney
% ...
% Ljubljana
% \author{N.~Zwahlen}\affiliation{\'Ecole Polytechnique F\'ed\'erale de Lausanne (EPFL), Lausanne 1015} % Lausanne
% \author{O.~Zyukova}\affiliation{Budker Institute of Nuclear Physics SB RAS and Novosibirsk State University, Novosibirsk 630090} % BINP
\collaboration{The Belle Collaboration}

\maketitle

{\renewcommand{\thefootnote}{\fnsymbol{footnote}}}
\setcounter{footnote}{0}

%%% Introduction
\section{Introduction}
\label{sec:introduction}

Properties of heavy-flavored hadrons such as masses and decay widths can, in principle, be described in the theoretical framework of quantum chromodynamics (QCD). However, they are difficult to calculate in practice with the perturbative QCD technique due to the fact that the strong coupling constant $\alpha_{s}$ is large in this low energy regime. To overcome this difficulty, other methods such as lattice QCD \protect\cite{Mathur01,Mathur02,Namekawa13}, heavy quark effective theory \protect\cite{Roberts08}, quark model \protect\cite{Ebert08}, QCD sum rule \protect\cite{Zhang08}, and bag model \protect\cite{Bernotas09} are deployed.

The properties of the $\Sigma_{c}^{0/++}$ baryons have been measured by many experiments \protect\cite{resultE791,resultFOCUSMass,resultFOCUSWidth,resultCLEO2455,resultCLEO2520,resultCDF,resultBaBar}, but the total uncertainties of the world averages remain large \protect\cite{PDG}. For example, the relative uncertainties of the decay widths are around 10\% of their central values. Furthermore, the relative uncertainty of the mass splitting $m(\Sigma_{c}(2455)^{++})-m(\Sigma_{c}(2455)^{0})$ is about 40\%, and there is no significant measurement for the mass splitting $m(\Sigma_{c}(2520)^{++})-m(\Sigma_{c}(2520)^{0})$ \protect\cite{resultCLEO2520,resultCLEO22520}. Due to the mass hierarchy between the $d$ and $u$ quarks, one may expect that the $\Sigma_{c}^{0}$ $(ddc)$ baryon is heavier than the $\Sigma_{c}^{++}$ $(uuc)$ baryon; however, many experimental results contradict this naive expectation \protect\cite{resultE791,resultCLEO2455,resultCLEO2520,resultCDF}. To explain the discrepancy, various models have been introduced \protect\cite{Chan85,Hwang87,Capstick87,Verma88,Cutkosky93,Genovese98,Silvestre03} that predict positive mass splittings. Precise measurements of the mass splittings are necessary to test these models.

In this paper, we present precise measurements of the masses and decay widths of the $\Sigma_{c}(2455)^{0/++}$ and $\Sigma_{c}(2520)^{0/++}$ baryons, and of their mass splittings. Throughout this paper, the charge-conjugate decay modes are implied. 

%%% Data samples and selection
\section{Data samples and event selections}
\label{sec:datasamples}

This study uses a data sample corresponding to an integrated luminosity of 711 fb$^{-1}$ collected with the Belle detector at the KEKB $e^{+}e^{-}$ asymmetric-energy collider \protect\cite{KEKB} operating at the $\Upsilon(4S)$ resonance. The Belle detector is a large solid angle magnetic spectrometer that consists of a silicon vertex detector (SVD), a 50-layer central drift chamber (CDC), an array of aerogel threshold Cherenkov counters (ACC), a barrel-like arrangement of time-of-flight scintillation counters (TOF), and an electromagnetic calorimeter comprising CsI(Tl) crystals located inside a superconducting solenoid coil that provides a 1.5 T magnetic field. An iron flux return located outside the coil is instrumented to detect $K^{0}_{L}$ mesons and to identify muons. A detailed description of the Belle detector can be found in Ref. \protect\cite{BelleDetector}.

\begin{figure*}[btp]
\centering 
\includegraphics[width=.49\textwidth]{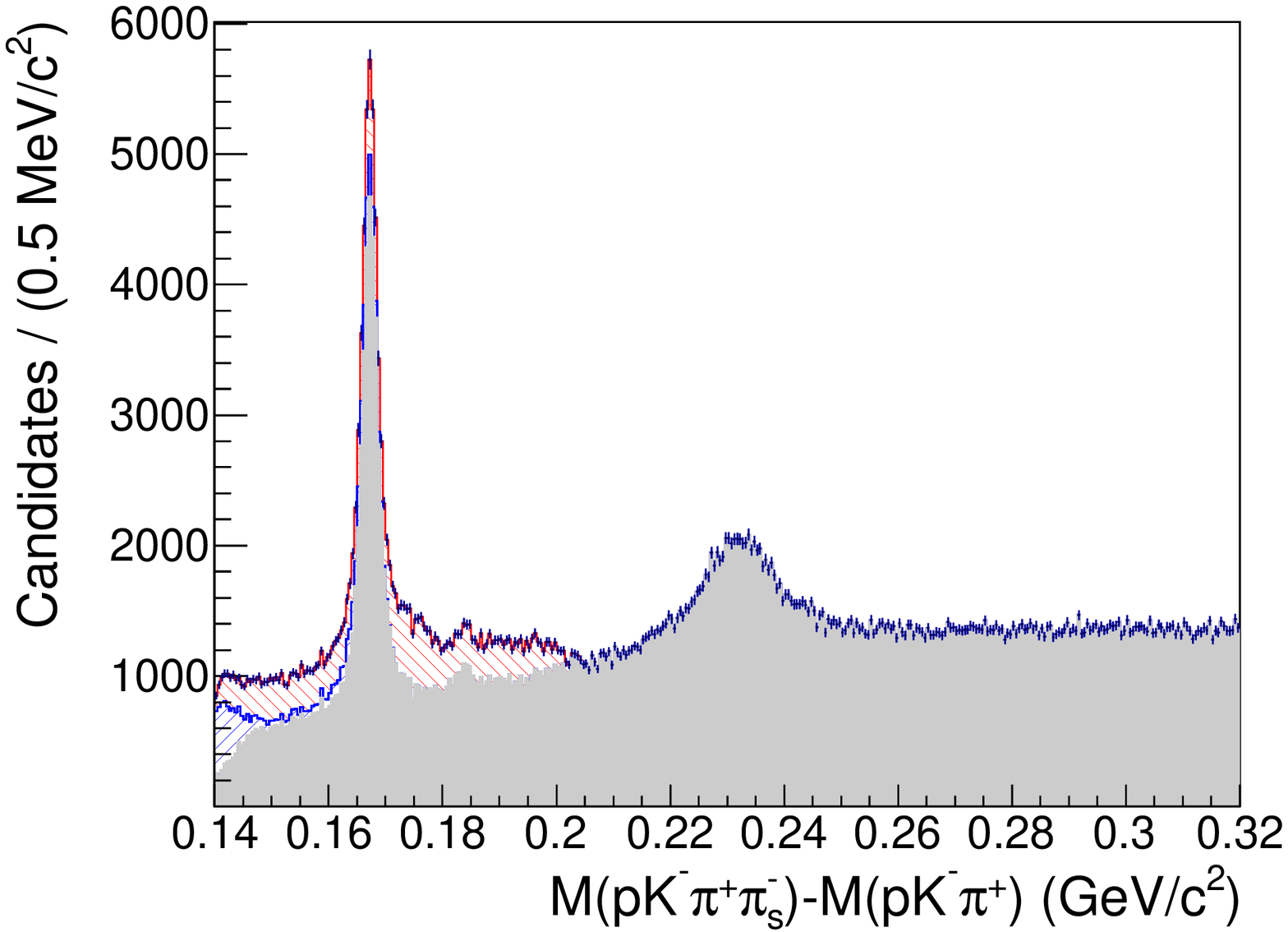}
\includegraphics[width=.49\textwidth]{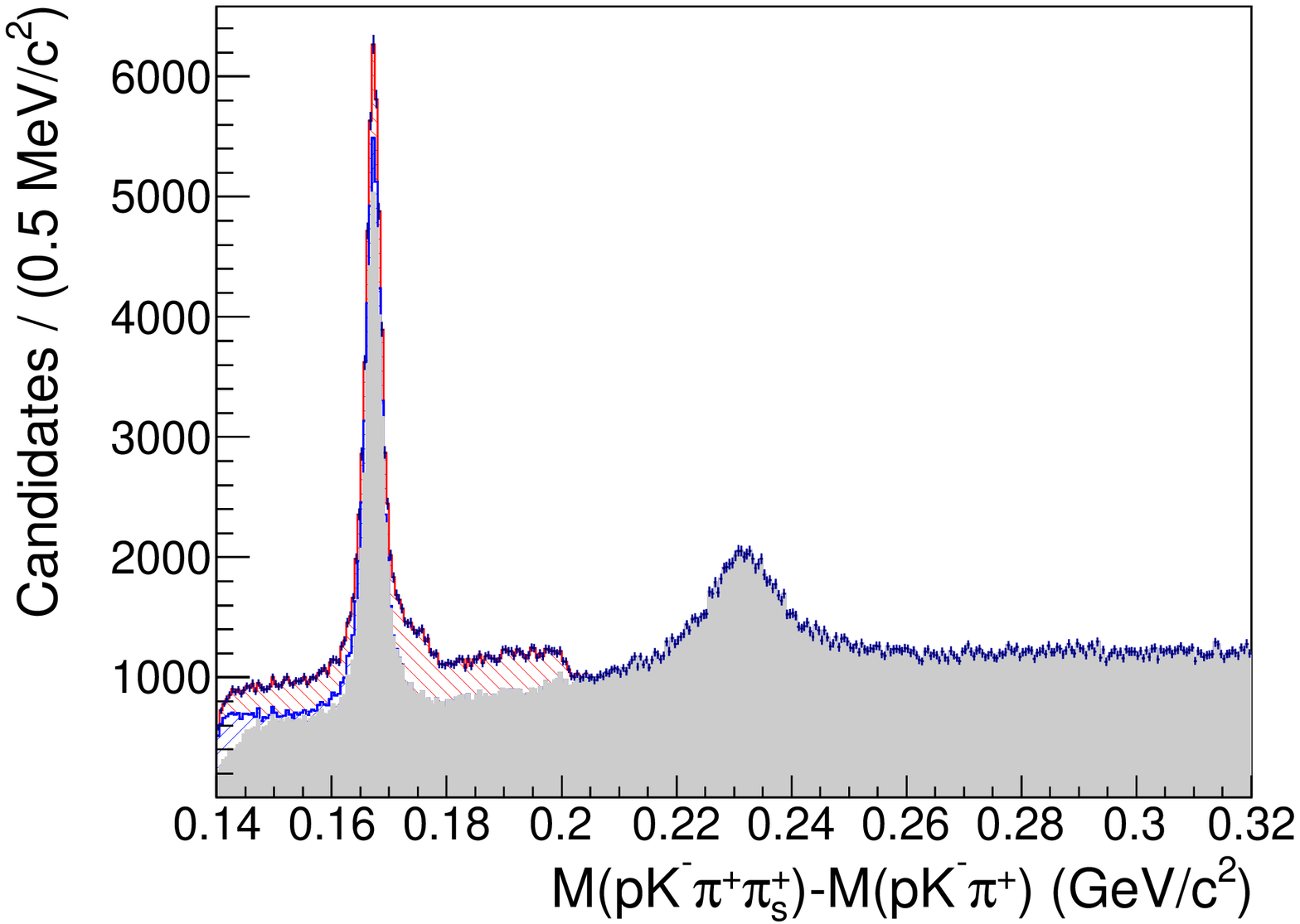}
\caption{\label{fig:feeddown} $M(pK^{-}\pi^{+}\pi^{-}_{s})-M(pK^{-}\pi^{+})$ (left) and $M(pK^{-}\pi^{+}\pi^{+}_{s})-M(pK^{-}\pi^{+})$ (right) distributions before (points) and after (shaded) the feed-down subtraction. The subtracted feed-down backgrounds from the $\Lambda_{c}(2595)^{+}$ (left-hatched) and $\Lambda_{c}(2625)^{+}$ (right-hatched) are also shown. The first and second peaks correspond to the $\Sigma_{c}(2455)^{0/++}$ and $\Sigma_{c}(2520)^{0/++}$ signals.}
\end{figure*}

The $\Sigma_{c}^{0/++}$ baryons are reconstructed via their $\Sigma_{c}^{0/++}\rightarrow\Lambda_{c}^{+}(\rightarrow pK^{-}\pi^{+})\pi^{-/+}_{s}$ decays, where $\pi_{s}$ is a low-momentum (``slow'') pion. Charged tracks are required to have an impact parameter with respect to the interaction point of less than 3 cm along the beam direction (the $z$ axis) and less than 1 cm in the plane transverse to the beam direction. In addition, each track is required to have at least two associated vertex detector hits each in the $z$ and azimuthal strips of the SVD. The particles are identified using likelihood \protect\cite{BellePID} criteria that have efficiencies of 84\%, 91\%, 93\%, and 99\% for $p$, $K$, $\pi$, and $\pi_{s}$, respectively. $\Lambda_{c}^{+}$ candidates are reconstructed as combinations of $p$, $K^{-}$, and $\pi^{+}$ candidates with an invariant mass between 2278.07 and 2295.27 MeV/$c^{2}$, corresponding to $\pm2.1\sigma$ around the nominal $\Lambda_{c}^{+}$ mass, where $\sigma$ represents the $\Lambda_{c}^{+}$ invariant mass resolution. $\Lambda_{c}^{+}$ daughter tracks are refit assuming they originate from a common vertex. The $\Lambda_{c}^{+}$ production vertex is defined by the intersection of its trajectory with the $e^{+}e^{-}$ interaction region. $\Lambda_{c}^{+}$ candidates are combined with $\pi_{s}$ candidates to form $\Sigma_{c}^{0/++}$ candidates. $\pi_{s}$ candidates are required to originate from the $\Lambda_{c}^{+}$ production vertex in order to improve their momentum resolution, which results in an enhanced signal-to-background ratio. Signal candidates retained for further analysis are required to have a confidence level greater than 0.1\% for the $\pi_{s}$ vertex fit constrained to the $\Lambda_{c}^{+}$ production vertex. To suppress combinatorial backgrounds, we also require the momentum of $\Sigma_{c}^{0/++}$ baryons in the center-of-mass frame to be greater than 2.0 GeV/$c$. The distributions of the mass difference $\Delta M\equiv M(pK^{-}\pi^{+}\pi^{-/+}_{s})-M(pK^{-}\pi^{+})$ for all reconstructed $\Sigma_{c}^{0/++}$ candidates are shown in Fig. \ref{fig:feeddown}.

We also use a Monte Carlo (MC) simulation sample for various purposes in this study, where events are generated with PYTHIA \protect\cite{PYTHIA}, decays of unstable particles are modeled with EVTGEN \protect\cite{EvtGen}, and the detector response is simulated with GEANT3 \protect\cite{GEANT}.

%%% Backgrounds
\section{Backgrounds}
\label{sec:backgrounds}

The sample of selected $\Sigma_{c}^{0/++}$ candidates includes two types of backgrounds: partially reconstructed decays of excited $\Lambda_{c}^{+}$ baryons (referred to as ``feed-down backgrounds'') and random combinations of the final state particles. The procedures used to parameterize these backgrounds are described in this section.

\subsection{Feed-down backgrounds from excited $\boldmath{\Lambda_{c}^{+}}$ baryons}
\label{sec:feeddownbackgrounds}

\begin{figure}[tbp]
\centering
\includegraphics[width=.49\textwidth]{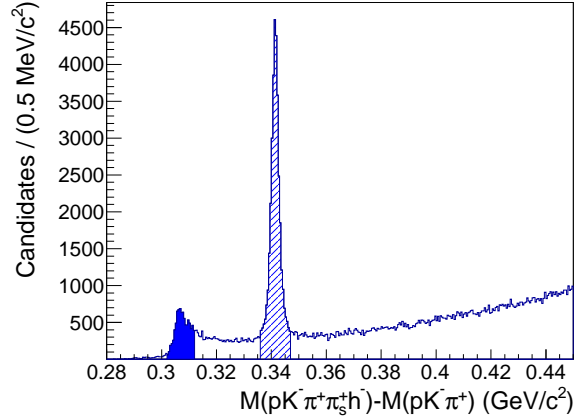}
\caption{\label{fig:excitedLambdac} Mass difference of $M(pK^{-}\pi^{+}\pi^{+}_{s}h^{-})-M(pK^{-}\pi^{+})$. Signal regions of the $\Lambda_{c}(2595)^{+}$ (filled) and $\Lambda_{c}(2625)^{+}$ (hatched) are defined in the text.}
\end{figure}

From the tracks of a $\Lambda_{c}^{*+}\rightarrow\Lambda_{c}^{+}\pi^{+}_{s}\pi^{-}_{s}$ decay, a $\Sigma_{c}$ candidate can be reconstructed if one of the slow pions is left out. This can be either a signal (from a $\Sigma_{c}^{0/++}$ resonant decay of an excited $\Lambda_{c}^{+}$ state) or a feed-down background event. The feed-down backgrounds from the $\Lambda_{c}(2595)^{+}$ and $\Lambda_{c}(2625)^{+}$ states appear in the $\Sigma_{c}(2455)^{0/++}$ mass region. In order to remove these backgrounds, we tag events that have a mass difference $M(pK^{-}\pi^{+}\pi_{s}^{-/+}h^{+/-})-M(pK^{-}\pi^{+})$ ($h^{+/-}$ being a charged track) that falls either in the [302, 312] MeV/$c^{2}$ or the [336, 347] MeV/$c^{2}$ mass interval, corresponding to the $\Lambda_{c}(2595)^{+}$ and $\Lambda_{c}(2625)^{+}$ signals, respectively (see Fig. \ref{fig:excitedLambdac}). The tagged events are subtracted from the $\Delta M$ distributions as shown in Fig. \ref{fig:feeddown}. To prevent a possible bias in the subtraction, we estimate the backgrounds under the $\Lambda_{c}^{*+}$ peaks from MC simulations and subtract them from the tagged feed-down backgrounds. Furthermore, we take into account the charged track detection efficiency of 74\% on average to correct for the feed-down backgrounds. Since the shape of the feed-down backgrounds depends on the $\pi_{s}$ momentum, we obtain and apply the efficiency correction as a function of this quantity. 

\subsection{Random backgrounds}
\label{sec:randombackgrounds}

The remaining background consists of random combinations, with or without a true $\Lambda_{c}^{+}$ baryon. In the latter case, the background level is estimated from the $\Lambda_{c}^{+}$ mass sidebands, defined as $M(pK^{-}\pi^{+})$ $\in$ [2259.16, 2267.76] MeV/$c^{2}$ or $M(pK^{-}\pi^{+})$ $\in$ [2305.58, 2314.18] MeV/$c^{2}$. The treatment of the random backgrounds in the fit is discussed in Sec. \ref{sec:fitprocedure}.

%%% Fit Procedure
\section{Fit procedure}
\label{sec:fitprocedure}

\begin{figure*}[tbp]
\centering 
\includegraphics[width=.49\textwidth]{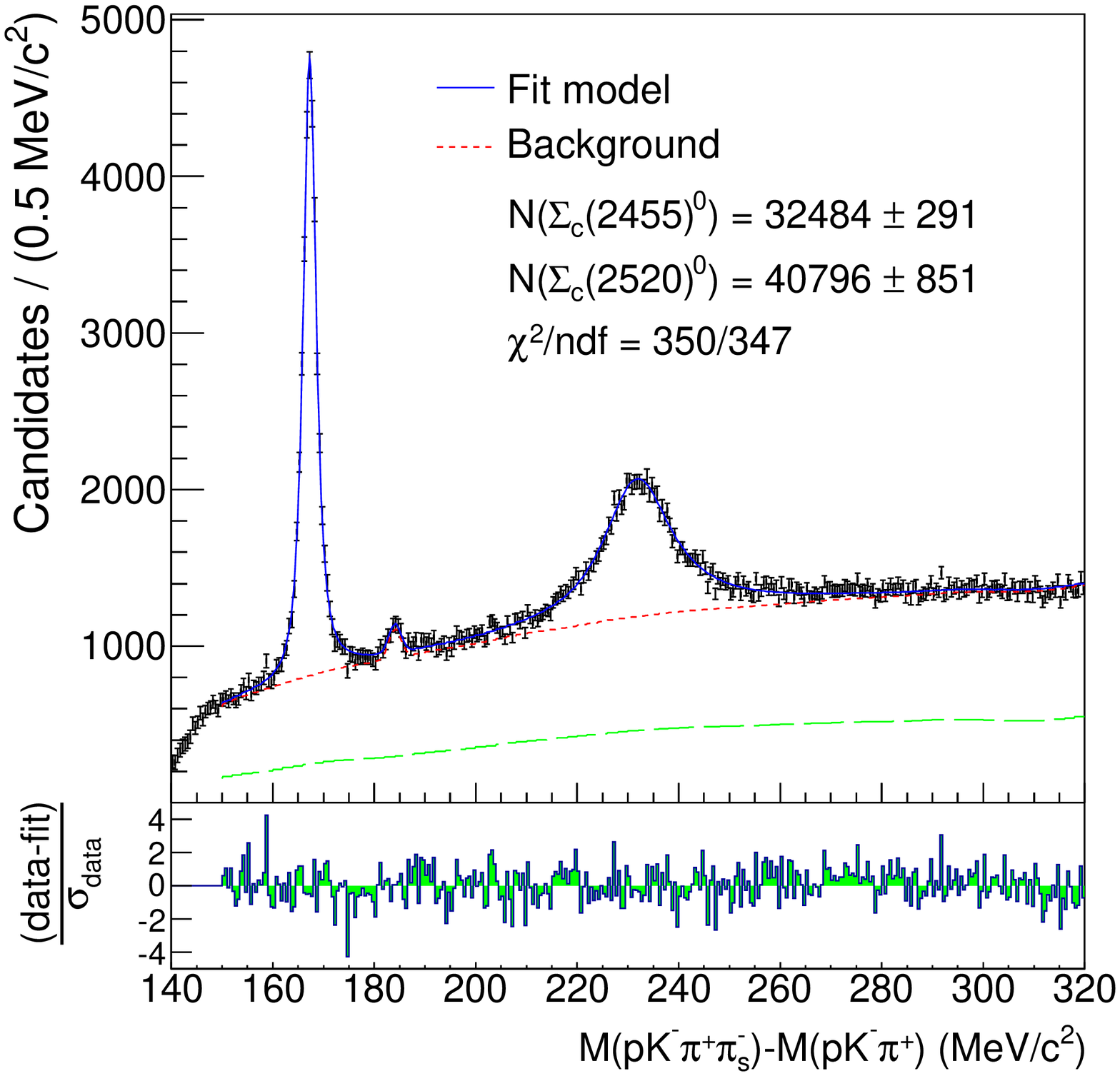}
\includegraphics[width=.49\textwidth]{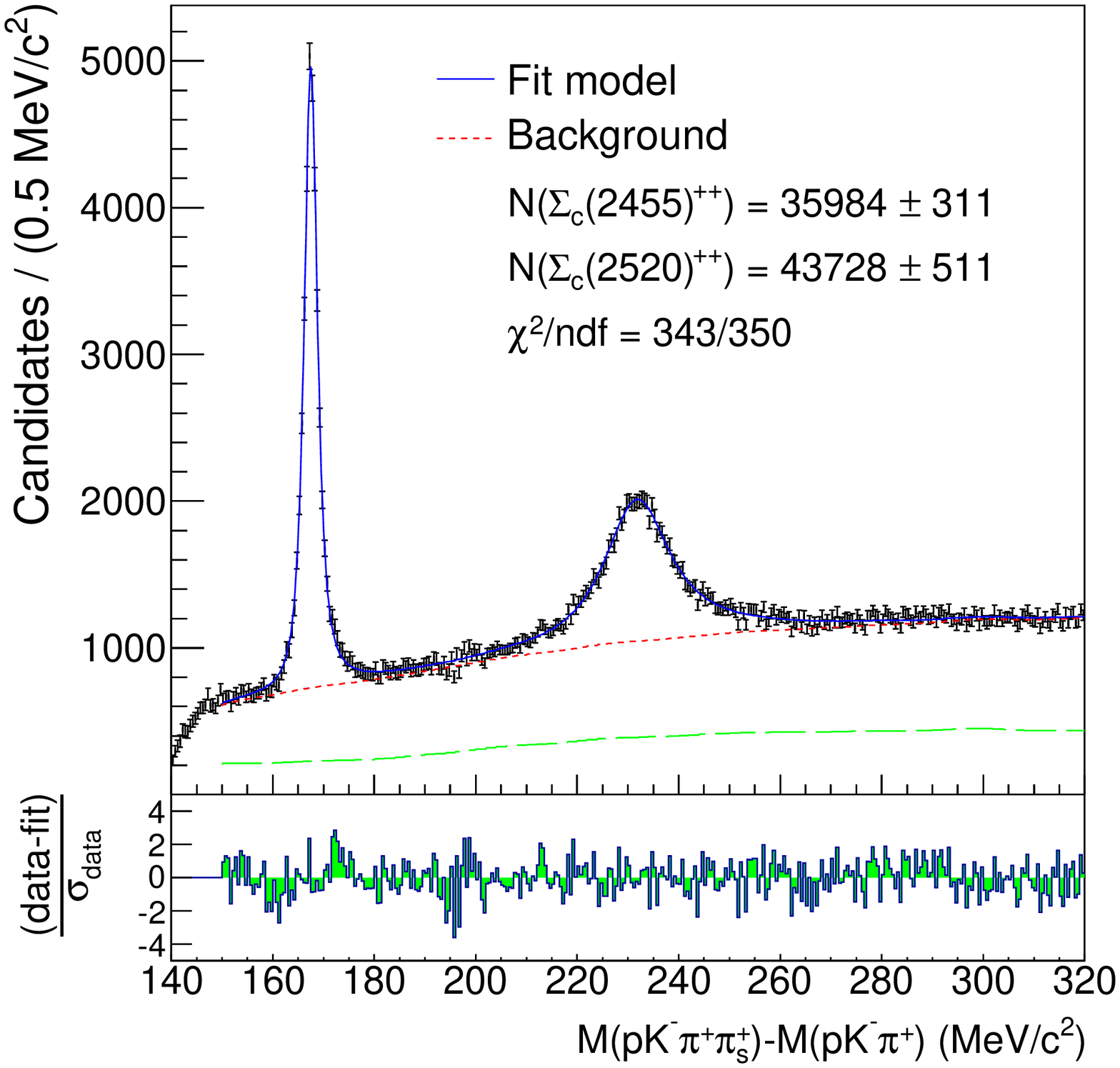}
\caption{\label{fig:globalfit} Fits to the mass differences $M(pK^{-}\pi^{+}\pi^{-}_{s})-M(pK^{-}\pi^{+})$ (left) and $M(pK^{-}\pi^{+}\pi^{+}_{s})-M(pK^{-}\pi^{+})$ (right) obtained from data (points with error bar) with the models (solid lines) described in the text. The random backgrounds without true $\Lambda_{c}^{+}$ baryons (long-dashed line) and the total backgrounds (dashed lines) are shown as well. The peak near 185 MeV/$c^{2}$ in the left plot is due to the $\Xi_{c}^{0}\rightarrow\Lambda_{c}^{+}\pi^{-}$ decay. The fit signal yields as well as the fit $\chi^{2}$ per degree of freedom are indicated on the plots. The bottom histograms are the differences between the values of data and fit divided by the statistical uncertainties of data to illustrate the fit quality.}
\end{figure*}

The parameters of the $\Sigma_{c}(2455)^{0/++}$ and $\Sigma_{c}(2520)^{0/++}$ signals, namely the decay widths and the mass differences with respect to the $\Lambda_{c}^{+}$ mass, are determined by performing binned maximum likelihood fits. Due to the small fraction of the weighted events in the region where the feed-down background is subtracted, a correction to the covariance matrix of the fit parameters is applied to obtain the proper errors. The $\Sigma_{c}(2455)^{0/++}$ and $\Sigma_{c}(2520)^{0/++}$ baryons are described by a relativistic Breit-Wigner probability density function (PDF) convolved with the detector response function as 
\begin{eqnarray*}
\int^{+\infty}_{-\infty}T(\Delta M';\Delta M_{0},\Gamma)R(\Delta M - \Delta M')d(\Delta M')
\end{eqnarray*}
where $T(\Delta M;\Delta M_{0},\Gamma)$ is a relativistic Breit-Wigner with the nominal mass difference $\Delta M_{0}\equiv M(\Sigma_{c})-M(\Lambda_{c}^{+})$ and the decay width $\Gamma$  as fit parameters, and $R$ is the detector response function. 

The resolution function $R$ is parameterized as the sum of three Gaussian functions centered at zero. The parameters are obtained from an MC simulation separately for the $\Sigma_{c}(2455)$ and $\Sigma_{c}(2520)$ signals. The detector resolutions for the $\Sigma_{c}(2455)$ and $\Sigma_{c}(2520)$ baryons are found to be $1.012\pm0.001$ and $1.578\pm0.013$ MeV/$c^{2}$, respectively, from the weighted variances of the three Gaussian distributions where the errors are statistical.

The random backgrounds without true $\Lambda_{c}^{+}$ baryons are modeled as histogram PDFs with shape and normalization taken from the $\Lambda_{c}^{+}$ baryon data sidebands. The random backgrounds with true $\Lambda_{c}^{+}$ baryons are described with a threshold function:
\begin{eqnarray*}
(\Delta M - m_{\pi})^{c_{0}}e^{c_{1}(\Delta M - m_{\pi})},
\end{eqnarray*}
where $c_{0}$, $c_{1}$ are fit parameters and $m_{\pi}$ is the known charged pion mass \protect\cite{PDG}.

In the neutral channel, we find a small peak near $\Delta M=185$ MeV/$c^{2}$. Based on studies performed using MC and data samples, we confirm the origin of this peak to be the as-of-yet unobserved decay of $\Xi_{c}^{0}\rightarrow\Lambda_{c}^{+}\pi^{-}$. We describe this peak with a Gaussian function. The mean and width of the Gaussian from the fit are found to be $184.08\pm0.15$ and $1.21\pm0.17$ MeV/$c^{2}$, respectively; the former is consistent with that from the world average ($m(\Xi_{c}^{0})-m(\Lambda_{c}^{+})=184.42^{+0.37}_{-0.81}$ MeV/$c^{2}$) \protect\cite{PDG} and the latter is consistent with that from MC.

The fit results to $\Delta M$ are shown in Fig. \ref{fig:globalfit}. The goodness-of-fit values are $\chi^{2}=350$ with 347 degrees of freedom for $\Sigma_{c}^{0}$ and $\chi^{2}=343$ with 350 degrees of freedom for $\Sigma_{c}^{++}$.

%%% Systematics
\section{Systematic uncertainties}
\label{sec:systematics}
To estimate systematic uncertainties, three sources are studied: momentum scale, resolution and fit model, and background parameterization. These are summarized in Table \ref{table:systematics}.

\subsection{Momentum calibration}
\label{sec:sysMomentumscale}

\begin{figure}[tbp]
\centering
\includegraphics[width=.49\textwidth]{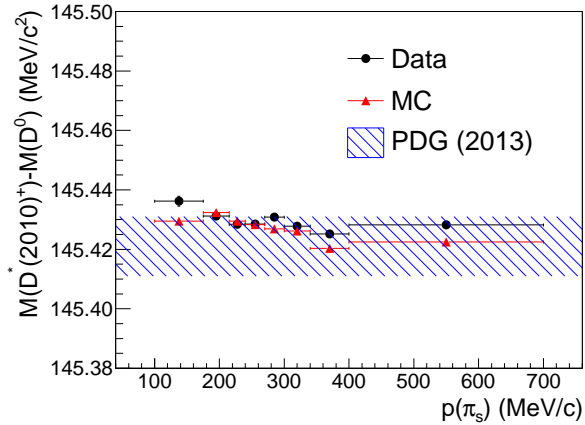}
\caption{\label{fig:systResolution} Mass difference $M(D^{*}(2010)^{+})-M(D^{0})$ obtained from MC (red triangle) and data (black circle) using the $D^{*}(2010)^{+}\rightarrow D^{0}(\rightarrow K^{-}\pi^{+})\pi^{+}_{s}$ decay as a function of the $\pi_{s}$ momentum. The uncertainties of each points are too small to be displayed. The world average with its total uncertainty \protect\cite{PDG} is also shown as a hatched area.}
\end{figure}

Mass measurements are sensitive to the momentum scale of the detector. Because there is a possible bias in the measurements of the charged track momenta, which may be due to the energy loss of the charged particles in materials, one should consider the precision of the momentum calibration. To minimize the possible bias, we calibrate the momentum scale using the copious $K^{0}_{S}\rightarrow\pi^{+}\pi^{-}$ sample. Charged tracks are iteratively calibrated as functions of the curvature, polar angle, and momentum of each track in the laboratory frame by comparing the reconstructed and world average \protect\cite{PDG} masses of $K^{0}_{S}$ meson as a function of the $K^{0}_{S}$ momentum. The obtained corrections are applied to the data sets used in this study. To estimate the accuracy, we choose a control sample of $D^{*}(2010)^{+}\rightarrow D^{0}(\rightarrow K^{-}\pi^{+})\pi^{+}_{s}$ decay, and compare the mass difference of $M(D^{*}(2010)^{+})-M(D^{0})$ over the $\pi_{s}$ momentum bins with the world average \protect\cite{PDG} as shown in Fig. \ref{fig:systResolution}. We observe the largest difference to be 0.02 MeV/$c^{2}$, which we assign as the systematic uncertainty on the mass difference measurements due to the momentum calibration.

\subsection{Resolution model}
\label{sec:systResolution}

Since our detector resolution model is evaluated from the MC as discussed in Sec. \ref{sec:fitprocedure}, the discrepancy between the MC and data is considered as a source of systematic uncertainty. To estimate the discrepancy, we compare the detector resolution in data and MC using the same control sample of $D^{*}(2010)^{+}\rightarrow D^{0}(\rightarrow K^{-}\pi^{+})\pi^{+}_{s}$ decay. Since the decay width of the $D^{*}(2010)^{+}$ meson is small, one can assume that the distribution of the mass difference $M(D^{*}(2010)^{+})-M(D^{0})$ is dominated by the detector resolution. We vary the widths of the detector response functions from +1.7\% to +11.8\% in the fits to $\Delta M$ by choosing the largest and smallest differences between the MC and data obtained by comparing $M(D^{*}(2010)^{+})-M(D^{0})$ as a function of the $\pi_{s}$ momentum. The uncertainties are found to be 0.19, 0.25, and 0.24 MeV/$c^{2}$ for the widths of the $\Sigma_{c}(2455)^{0/++}$,  $\Sigma_{c}(2520)^{0}$, and $\Sigma_{c}(2520)^{++}$ baryons, respectively. We also vary the detector response functions by $\pm 1\sigma$ deviation from the fitted resolution parameters, where $\sigma$ is the statistical error, and only small uncertainties are found for the decay widths of 0.01 and 0.04 MeV/$c^{2}$ for the $\Sigma_{c}(2455)^{0/++}$ and $\Sigma_{c}(2520)^{0/++}$ baryons, respectively.

\subsection{Fit model}
\label{sec:sysFitmodel}

We also check the internal consistency of the fitting procedure. In order to probe any bias from the fitter, we perform 10,000 pseudo-experiments for each of the mass differences, $\Delta M_{0}(\Sigma_{c}(2455))$ and $\Delta M_{0}(\Sigma_{c}(2520))$, and the decay widths $\Gamma(\Sigma_{c}(2455))$ and $\Gamma(\Sigma_{c}(2520))$. In the production of the pseudo-experiments, we set the input values to be those obtained from the data. From the study, we find negligible discrepancies.

The effect of binning is studied by varying the bin size in the fits to $\Delta M$ from 0.1 MeV/$c^{2}$ to 1.0 MeV/$c^{2}$. The uncertainties of $\Delta M_{0}$ are negligible, and we find small uncertainties for the widths of 0.09, 0.06, 0.04, and 0.05 MeV/$c^{2}$ for the $\Sigma_{c}(2455)^{0}$, $\Sigma_{c}(2455)^{++}$, $\Sigma_{c}(2520)^{0}$, and the $\Sigma_{c}(2520)^{++}$ baryons, respectively.

We also test the effect of various fit ranges. We choose several fit ranges, some of which include both the $\Sigma_{c}(2455)^{0/++}$ and $\Sigma_{c}(2520)^{0/++}$ signals and others only one of them. Though the results from the various fit ranges are consistent within the statistical fluctuations, we conservatively assign the variations in the fit results, 0.03 and 0.01 MeV/$c^{2}$ for $\Delta M_{0}(\Sigma_{c}(2520)^{0})$ and $\Delta M_{0}(\Sigma_{c}(2520)^{++})$, respectively, and 0.19 and 0.17 MeV/$c^{2}$ for the widths of the $\Sigma_{c}(2520)^{0}$ and $\Sigma_{c}(2520)^{++}$ baryons, respectively, as systematic uncertainties.

\begin{table*}[btp]
\centering
\caption{\label{table:systematics} Systematic uncertainties for the mass differences ($\Delta M_{0}$) and the decay widths ($\Gamma$) of the $\Sigma_{c}(2455)^{0/++}$ and $\Sigma_{c}(2520)^{0/++}$ baryons in MeV/$c^{2}$. The uncertainties for $\Delta M_{0}$ from the resolution model and for $\Gamma$ from the momentum calibration are insignificant.}
\begin{tabular}{c|cc|cc|cc|cc}
\hline\hline
& \multicolumn{2}{c|}{$\Sigma_{c}(2455)^{0}$} & \multicolumn{2}{c|}{$\Sigma_{c}(2520)^{0}$} & \multicolumn{2}{c|}{$\Sigma_{c}(2455)^{++}$} & \multicolumn{2}{c}{$\Sigma_{c}(2520)^{++}$} \\
\cline{2-9}
& $\Delta M_{0}$ & $\Gamma$ & $\Delta M_{0}$ & $\Gamma$ & $\Delta M_{0}$ & $\Gamma$ & $\Delta M_{0}$ & $\Gamma$ \\
\hline
Momentum calibration & $\pm0.02$ & -- & $\pm0.02$ & -- & $\pm0.02$ & -- & $\pm0.02$ & -- \\
~~~~~Resolution model~~~~~ & -- & ${}^{+0.01}_{-0.19}$ & -- & ${}^{+0.04}_{-0.25}$ & -- & ${}^{+0.01}_{-0.19}$ & -- & ${}^{+0.04}_{-0.24}$ \\
Fit model & ~$\pm0.01$~ & ~$\pm0.09$~ & ~$\pm 0.03$~ & ~$\pm0.20$~ & ~$\pm0.01$~ & ~$\pm0.07$~ & ~$\pm 0.01$~ & ~$\pm0.18$~ \\
\hline
Total & $\pm0.02$ & ${}^{+0.09}_{-0.21}$ & $\pm0.04$ & ${}^{+0.20}_{-0.32}$ & $\pm0.02$ & ${}^{+0.07}_{-0.20}$ & $\pm0.02$ & ${}^{+0.18}_{-0.30}$ \\
\hline\hline
\end{tabular}
\caption{\label{table:result} The measurements of the masses ($M_{0}$) and the widths ($\Gamma$) of the $\Sigma_{c}(2455)^{0/++}$ and $\Sigma_{c}(2520)^{0/++}$ baryons. The first error is statistical and the second is systematic. The masses are calculated by adding the world average of $\Lambda_{c}^{+}$ mass to the mass differences  ($\Delta M_{0}$) and the third error is the total uncertainty of the world average of $\Lambda_{c}^{+}$ mass \protect\cite{PDG}.}
\begin{tabular}{l|ccc}
\hline\hline
 & $\Delta M_{0}$ (MeV/$c^{2}$) & $\Gamma$ (MeV/$c^{2}$)  & $M_{0}$ (MeV/$c^{2}$) \\
\hline
~$\Sigma_{c}(2455)^{0}$~ & ~~$167.29 \pm 0.01 \pm 0.02$~~ & $\phantom{1}1.76 \pm 0.04^{+0.09}_{-0.21}$ & ~~~$2453.75 \pm 0.01 \pm 0.02 \pm 0.14$~~ \\
~$\Sigma_{c}(2455)^{++}$~ & $167.51 \pm 0.01 \pm 0.02$ & $\phantom{1}1.84 \pm 0.04^{+0.07}_{-0.20}$ & ~~~$2453.97 \pm 0.01 \pm 0.02 \pm 0.14$~~ \\
~$\Sigma_{c}(2520)^{0}$~ & ~~$231.98 \pm 0.11 \pm 0.04$~~ & ~~$15.41 \pm 0.41^{+0.20}_{-0.32}$~~ & ~~$2518.44 \pm 0.11 \pm 0.04 \pm 0.14$~~\\
~$\Sigma_{c}(2520)^{++}$~ & $231.99 \pm 0.10 \pm 0.02$ & $14.77 \pm 0.25^{+0.18}_{-0.30}$ & ~~$2518.45 \pm 0.10 \pm 0.02 \pm 0.14$~~\\ 
\hline\hline
\end{tabular}
\end{table*}

\subsection{Background model}
\label{sec:sysBackgroundmodel}

Since we correct the feed-down backgrounds by taking into account the efficiency as discussed in Sec. \ref{sec:backgrounds}, the uncertainty of the efficiency should also be taken into account. The systematic uncertainty from the feed-down model is estimated as 1.87\% from the error propagation of the statistical uncertainties of the feed-down backgrounds, the uncertainties of the tracking efficiency and the acceptance of the detector. We vary the yields of the feed-down background by $\pm1.87\%$ without significant effect on the fit results compared with the statistical uncertainties. Since we fix the yields of the random backgrounds without true $\Lambda_{c}^{+}$ baryons, as discussed in Sec. \ref{sec:backgrounds}, we also vary the yields of the random backgrounds by their uncertainties; only negligible effects are obtained. Finally, we test other threshold functions to describe the random backgrounds with true $\Lambda_{c}^{+}$ baryons, but again find only negligible effects.

%%% Results
\section{Results}
\label{sec:result}
Our measurements for the mass differences (with respect to the $\Lambda_{c}^{+}$ mass) and the decay widths of the $\Sigma_{c}(2455)^{0/++}$ and $\Sigma_{c}(2520)^{0/++}$ baryons are summarized in Table \ref{table:result}. We also calculate the mass splittings $M_{0}(\Sigma_{c}^{++})-M_{0}(\Sigma_{c}^{0})$ from $\Delta M_{0}(\Sigma_{c}^{0})$ and $\Delta M_{0}(\Sigma_{c}^{++})$ as $M_{0}(\Sigma_{c}(2455)^{++})-M_{0}(\Sigma_{c}(2455)^{0})=0.22\pm0.01\pm0.01$ MeV/$c^{2}$ and $M_{0}(\Sigma_{c}(2520)^{++})-M_{0}(\Sigma_{c}(2520)^{0})=0.01\pm0.15\pm0.03$ MeV/$c^{2}$ where the first error is statistical and the second is systematic. Since the mass splittings are calculated from $\Delta M_{0}$, most of the systematic uncertainties cancel, such as that from the momentum calibration. These measurements are the most precise to date. The mass splitting $M_{0}(\Sigma_{c}(2455)^{++})-M_{0}(\Sigma_{c}(2455)^{0})$ is found to be positive as expected by the models \protect\cite{Chan85,Hwang87,Capstick87,Verma88,Cutkosky93,Genovese98,Silvestre03}.

%***** Acknowledgments *****
%----------- Long version, for most papers ----------- 
\acknowledgments
We thank the KEKB group for the excellent operation of the
accelerator; the KEK cryogenics group for the efficient
operation of the solenoid; and the KEK computer group,
the National Institute of Informatics, and the 
PNNL/EMSL computing group for valuable computing
and SINET4 network support.  We acknowledge support from
the Ministry of Education, Culture, Sports, Science, and
Technology (MEXT) of Japan, the Japan Society for the 
Promotion of Science (JSPS), and the Tau-Lepton Physics 
Research Center of Nagoya University; 
the Australian Research Council and the Australian 
Department of Industry, Innovation, Science and Research;
Austrian Science Fund under Grant No. P 22742-N16;
the National Natural Science Foundation of China under Contracts 
No.~10575109, No.~10775142, No.~10825524, No.~10875115, No.~10935008 
and No.~11175187; 
the Ministry of Education, Youth and Sports of the Czech
Republic under Contract No.~LG14034;
the Carl Zeiss Foundation, the Deutsche Forschungsgemeinschaft
and the VolkswagenStiftung;
the Department of Science and Technology of India; 
the Istituto Nazionale di Fisica Nucleare of Italy; 
the WCU program of the Ministry Education Science and
Technology, National Research Foundation of Korea Grants
No.~2011-0029457, No.~2012-0008143, No.~2012R1A1A2008330,
No.~2013R1A1A3007772;
the BRL program under NRF Grant No.~KRF-2011-0020333,
No.~KRF-2011-0021196,
Center for Korean J-PARC Users, No.~NRF-2013K1A3A7A06056592; the BK21
Plus program and the GSDC of the Korea Institute of Science and
Technology Information;
the Polish Ministry of Science and Higher Education and 
the National Science Center;
the Ministry of Education and Science of the Russian
Federation and the Russian Federal Agency for Atomic Energy;
the Slovenian Research Agency;
the Basque Foundation for Science (IKERBASQUE) and the UPV/EHU under 
program UFI 11/55;
the Swiss National Science Foundation; the National Science Council
and the Ministry of Education of Taiwan; and the U.S.\
Department of Energy and the National Science Foundation.
This work is supported by a Grant-in-Aid from MEXT for 
Science Research in a Priority Area (``New Development of 
Flavor Physics'') and from JSPS for Creative Scientific 
Research (``Evolution of Tau-lepton Physics''). E. Won acknowledges support by NRF Grant No. 2010-0021174, B. R. Ko by NRF grant No. 2010-0021279.

\end{document}